\documentclass{article}

\usepackage{arxiv}

\usepackage[utf8]{inputenc} 
\usepackage[T1]{fontenc}    
\usepackage[hidelinks]{hyperref}
\usepackage{url}            
\usepackage{booktabs}       
\usepackage{amsfonts}       
\usepackage{nicefrac}       
\usepackage{microtype}      
\usepackage{lipsum}		

\usepackage{amsmath,amssymb}

\usepackage{graphicx}

\usepackage{caption}

\usepackage{cellspace}
\DeclareCaptionLabelFormat{myformat}{\textbf{#1 #2}} 
\DeclareUnicodeCharacter{2212}{-} 

\title{Information bounds on the accuracy of cell polarization}

\author{
    {Tau-Mu Yi}\\
	Department of Molecular, Cellular, and Developmental Biology\\
	University of California\\
	Santa Barbara, CA 93106 \\
	\texttt{tmy@ucsb.edu} \\
}

\renewcommand{\shorttitle}{Information bounds on the accuracy of cell polarization}

\hypersetup{
pdftitle={Information bounds on the accuracy of cell polarization},
pdfsubject={q-bio.NC, q-bio.QM},
pdfauthor={Tau-Mu Yi},
pdfkeywords={Information theory, Cell polarization, Gradient sensing},
}

\begin{document}
\maketitle

\begin{abstract}
Here we characterized an information measure for cell polarity that applies to non-motile cells responding to a chemical gradient. The central idea is that polarization represents information about the direction of the gradient. We applied a theory of optimal gradient sensing and response in the presence of external noise based on the information capacity of a Gaussian channel. First, we formulated an information framework that describes spatial gradient sensing and polarization response. As part of this section, we modeled ligand diffusion and receptor-binding dynamics as a mixed Poisson distribution, confirming the single receptor accuracy limits derived by ten Wolde and colleagues. Second, we performed numerical calculations of stochastic ligand levels at the cell surface to estimate the information provided about the directional component of the gradient vector, which was close to the Gaussian channel bound for low signal-to-noise ratios. Third, we used the information framework to evaluate the noise-robustness of three generic models of cell polarity, demonstrating that a filter-amplifier architecture and time integration can attenuate the detrimental impact of noise on polarity so that the model can approach the theoretical limits. Fourth, we compared the theory to published experimental data on yeast mating projection growth in a pheromone gradient, identifying the ligand association rate and integration time as two key parameters affecting directional accuracy. By varying these parameters, we showed that for certain ranges the theory is roughly in agreement with the data, and that the slow binding rate constant is a key limiting factor. We concluded that temporal averaging can help overcome the slow binding rate to achieve greater accuracy, but with the drawback of a slow mating response.
\end{abstract}


\section{Introduction}
Gradient sensing and response is a basic cellular behavior~\cite{levchenko_models_2002,bagorda_eukaryotic_2008}. Cells sense a chemical gradient and then respond by moving or projecting up the gradient. During this process, protein components localize to the front (or back) of the cell; this asymmetry induced by an external (e.g. gradient) or internal cue is referred to as cell polarity~\cite{drubin_origins_1996,mogilner_cell_2012,campanale_development_2017}. Specialized structures, functions, and behaviors develop from this asymmetry. Examples include neurons forming axons and dendrites, and neutrophils actively tracking foreign invaders such as bacteria~\cite{petri_neutrophil_2018,stoeckli_understanding_2018}.

There are two basic strategies to traverse a chemical gradient: temporal sensing and spatial sensing~\cite{tan_computational_2018}. In temporal sensing, the cell measures the change in chemical concentration as a function of time ($dC/dt$), and then decides whether to continue moving in the same direction or to change directions~\cite{macnab_gradient-sensing_1972,block_impulse_1982}. Here we focus on spatial sensing in which the cell measures the concentration differences on the cell surface ($dC/dx$), and based on this direct gradient information, moves or projects in the appropriate direction.

The response can be measured in terms of polarization, e.g. projection growth (chemotropism), or movement (chemotaxis). Typically, the directional accuracy of the response is assessed by calculating the angle ($\theta$) of the polarization/growth/movement with respect to the gradient direction~\cite{wilkinson_assays_1998,zigmond_chemotaxis_1998}. For nonmotile cells, one calculates the cosine of this angle between the projection and gradient ($cos(\theta)$)~\cite{segall_polarization_1993}. For motile cells, one measures the equivalent chemotaxis index (CI) which is the distance traveled up the gradient divided by the total distance~\cite{fuller_external_2010}.

It has been found that cells are quite remarkable at gradient sensing. For example, yeast, neurons, neutrophils and \textit{Dictyostelium} cells can sense shallow gradients and respond by projecting or moving up the gradient with good accuracy~\cite{zigmond_chemotaxis_1998,moore_robust_2008,devreotes_chemotaxis_1988,bhattacharjee_large-scale_2017,levchenko_models_2002}. However, this gradient sensing behavior is not perfect. Not all cells are perfectly aligned with the gradient, nor do they move perfectly straight up the gradient.

What are the limits to the accuracy of gradient sensing and response? These limits are imposed by external or internal noise~\cite{ten_wolde_fundamental_2016} to the cell; in the absence of noise, the sensing and response can be perfectly aligned with the gradient direction. Substantial progress has been made understanding how accurately cells can measure chemical concentrations~\cite{ten_wolde_fundamental_2016,mora_limits_2010,endres_accuracy_2008,bialek_physical_2005,hu_physical_2010}. In particular, researchers have characterized the constraints placed by external noise from diffusion and stochastic receptor binding. Berg and Purcell~\cite{berg_physics_1977} originally formulated the theory for noise arising from individual ligand molecules randomly diffusing into the neighborhood of receptors. Subsequent work has described the additional variance from stochastic binding of those molecules to receptors on the cell surface giving rise to uncertainty in the estimated concentration~\cite{kaizu_berg-purcell_2014,bialek_physical_2005}. 

Recently, ten Wolde and colleagues~\cite{ten_wolde_fundamental_2016} reviewed the literature on the limits to cellular sensing of chemicals at low concentrations. One organizing principle was the concept of receptor correlation time ($t_c$), i.e. "the timescale over which fluctuations in the state of the receptor, arising from the stochastic receptor-ligand binding, decay." Cells sense chemical ligands by measuring receptor occupancy over a time window $T$, and the variance of this time-averaged occupancy ($n_T$) depends on a factor of $T/(2t_c)$, which can be interpreted as the number of independent measurements. The review then characterized the various results in the literature as different approaches to estimating $t_c$.

Kaizu et al.~\cite{kaizu_berg-purcell_2014} took advantage of the formalism of Agmon and Szabo~\cite{agmon_theory_1990} for reversible diffusion-influenced reactions to calculate the zero-frequency limit of the correlation function from which the receptor correlation time $t_c$ was obtained in terms of renormalized association and dissociation rate constants. The end result was an expression for the fractional variance of the concentration estimate by a single receptor which consisted of two terms. The first represented the uncertainty or noise from diffusion and was identical to the expression of Berg and Purcell~\cite{berg_physics_1977}, and the second represented noise from receptor binding, which was absent in the original work.

Bialek and Setayeshgar~\cite{bialek_physical_2005} applied the fluctuation-dissipation theorem (FDT) to the differential equation system describing receptor binding. Fluctuations in both diffusion and receptor-ligand interaction "impede" the force of receptor binding, thereby relating spontaneous fluctuations in receptor occupancy to the linear response of receptor occupancy to changes in binding free energy. From the power spectrum of these fluctuations, one can calculate the receptor correlation time ($t_c$), and hence the variance in monitored ligand concentration. The result was very similar to the expression of Kaizu et al.~\cite{kaizu_berg-purcell_2014} with an identical receptor-binding noise term and a slightly different diffusion term.

Whereas the above work was concerned with the cell measuring the concentration of a chemical ligand at a given point or region in space, other research has addressed the question  of spatial sensing, i.e. determining gradient direction by monitoring chemical concentration at different points on the cell surface. What are the limits of spatial sensing? Endres and Wingreen~\cite{endres_accuracy_2008} studied spatial sensing taking into account noise arising from diffusion (but not receptor binding). They calculated a maximum likelihood estimate of the variance in the gradient vector by fitting to the measurements made by receptors on the cell surface, and then estimated the variance of the fit. Interestingly, the result was numerically identical to the concentration measurement uncertainty from diffusion for a perfectly absorbing sphere.

From an information theory perspective, Andrews and Iglesias~\cite{andrews_information-theoretic_2007} applied rate distortion theory to investigate the quantitative relationship between input-output mutual information and chemotactic accuracy. More specifically, they employed a numerical procedure (Blahut-Arimoto algorithm~\cite{cover_elements_2006}) to compute the minimum amount of mutual information needed between the gradient and response directions to achieve a certain chemotaxis index. They then were able to model various chemotactic strategies to assess optimality and to compare to experimental results in the organism \textit{Dictyostelium}.

Hu et al.~\cite{hu_quantifying_2011} employed an analytic approach to derive an estimate of the mutual information between the gradient direction and the spatial distribution of bound receptors. More specifically, they constructed a sufficient statistic $Z$ of receptor occupancy on the cell surface and then calculated the mutual information of the gradient direction with respect to $Z$. The result was a monotone function of the signal-to-noise ratio in which signal was a function of gradient slope and the noise was approximated as Gaussian. The authors used this mutual information estimate at the level of receptor-ligand binding as a benchmark to compare with the downstream chemotactic performance of the cell.

Here we explored the limits of spatial gradient sensing and polarization response using yeast as an example and extending previous work. Put simply, polarization represents information about gradient direction, with the mutual information being equal to the decrease in polarization entropy from the unpolarized state. We used coding theory and the capacity of a Gaussian channel to specify the limits of polarization in the presence of external noise. Numerical calculations estimated the directional component of the gradient information and demonstrated consistency with the theoretical bounds. Simulations using previously published generic models of cell polarization demonstrated noise-robustness can be improved via a two-stage architecture and time integration. Finally, we compared the theoretical bounds to published gradient sensing experiments in yeast, and found that projection accuracy is limited by a slow association rate constant which necessitates a slow polarization response to take advantage of time integration.

\section{Results}
\subsection{Cell polarity provides information about the gradient direction}
\subsubsection{Polarization entropy and information}
\begin{figure}[!h]
\includegraphics[width=1.0\textwidth]{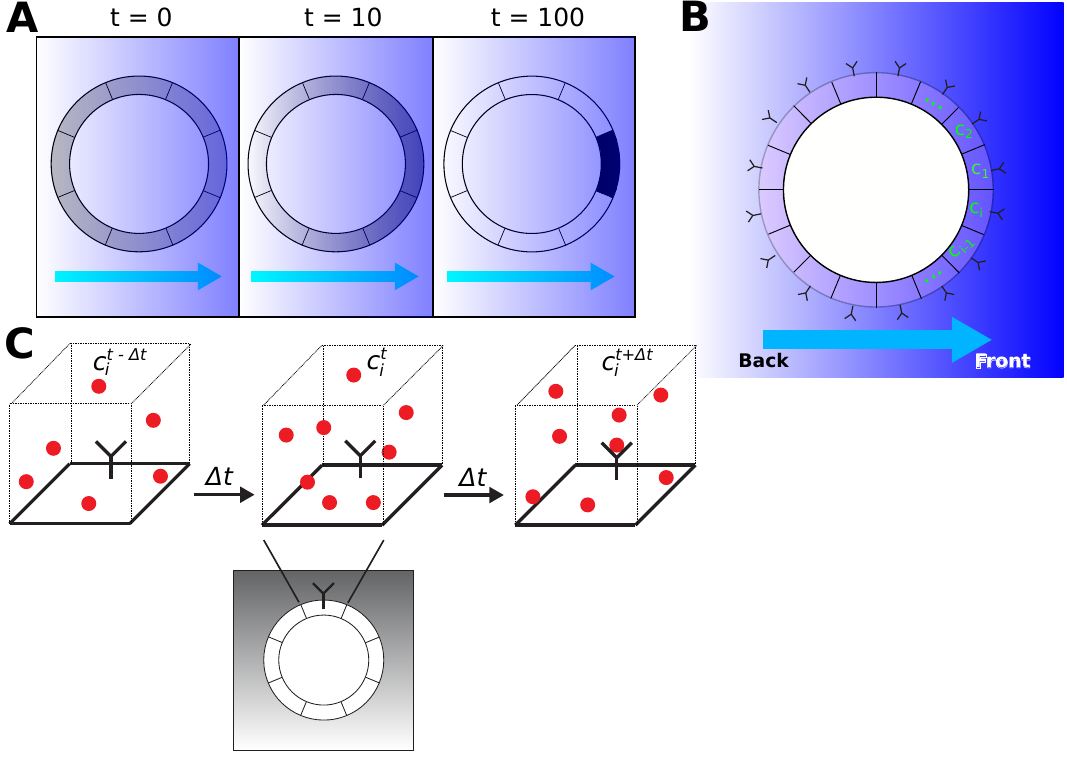} 
\caption{{\bf Schematic diagrams of gradient-directed cell polarization.}
A: Circular cell with the surface membrane divided into bins. The background shading and arrow underneath depict the gradient direction. A cellular species on the surface localizes to the front of the cell over a time interval resulting in polarization. B: Gradient sensing by receptors on the cell surface. Each cell surface compartment contains a single receptor (Y-shape). The receptor senses the gradient chemical concentration $c_i$ at bin $i$. C: The receptor measures the local concentration ($c_i$) of ligand (red circle) by binding ligand molecules over a time interval $\Delta t$. This monitoring process is subject to noise from the stochastic nature of diffusion (ligand molecules diffusing into and out of the receptor local neighborhood) and receptor-ligand binding.}
\label{fig1}
\end{figure}

In biological systems, cell polarity can be defined as the non-isotropic spatial distribution of cellular species with respect to a reference cue~\cite{drubin_origins_1996,campanale_development_2017}. Here we focus on directional polarization of a species on the cell surface (i.e. plasma membrane) in response to an external chemical gradient. We start with the concept of cell polarity as information about the gradient direction~\cite{hu_quantifying_2011}. For simplicity we represent the cell as a two-dimensional (2D) disk. The cell surface is divided into bins into which a polarized species can reside (Fig~\ref{fig1}A). 
The entropy of this distribution can be written as the polarization entropy $S_p$:
\begin{equation}
\label{eq:entropy}
S_p = -\sum_i^{n_b} p_i \log p_i
\end{equation}
in which $p_i$ represents the concentration of the polarized species (normalized to 1), $i$ is the bin index, $n_b$ is the number of bins, and the $\log$ function in this paper is base 2 so that entropy and information are in bits. Bins (compartments) of arbitrary size can be defined, and one possibility is that each bin contains a single receptor (Fig~\ref{fig1}B) that senses the local concentration of ligand (Fig~\ref{fig1}C) in the chemical gradient. 

Perfect polarization is when the polarization direction is correctly aligned to the reference (e.g. gradient) direction, and all of the species are located in the appropriate bin: $S_{p'} = 0$. The polarization direction is the vector from the center of the cell to the center of the polarization distribution on the membrane. Random polarization is when the species is unpolarized and equally distributed among all of the bins: $S_u = \log n_b$. 

The polarization entropy decreases as the cell goes from unpolarized (symmetric) to polarized (asymmetric) representing a gain in information. We define the information of polarization for a given polarized state $p$ as
\begin{equation}
\label{eq:info}
I_p = S_{u} - S_{p}
\end{equation}
which can be thought of as the mutual information $I(Y;X) = H(Y) - H(Y;X) = H(X) - H(X;Y) = I(X;Y)$ that the polarization $Y$ provides about the gradient $X$ (i.e. its direction) and vice versa (i.e. the gradient guides the direction of the polarization). We see that $I_p = I(Y;X)$, because $S_u = H(Y)$ (polarization entropy in absence of gradient), and $S_p = H(Y;X)$ (polarization entropy in presence of gradient).

This definition applies to a population of $m$ cells in which we average over the polarization profiles of the individual cells to calculate the population polarization entropy:
$S^{\text{pop}}_p = -\sum_i^{n_b} \bar{p_i} \log \bar{p_i}$ where $\bar{p_i} = \frac{\sum_j^m p_{ij}}{m}$ and the index $j$ corresponds to the $m$ cells with $p_{ij} = p_i$ for the $j$-th cell. 

\subsubsection{Defining polarization magnitude}

Polarization possesses a magnitude along with a direction, and the magnitude describes the extent of asymmetry. A cell is more polarized if the localized species is concentrated in a smaller region. For example, Lawson et al.~\cite{lawson_spatial_2013} quantified polarization as the full width at half maximum (FWHM) of the polarization profile of a species with a narrower width corresponding to greater polarization. 

\begin{figure}[!h]
\includegraphics[width=1.0\textwidth]{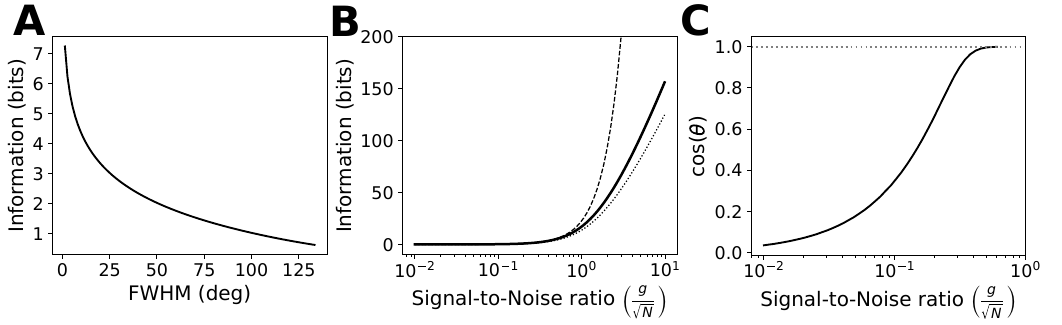} 
\caption{{\bf Relationships between different measures of polarization on 2D disk.}
A: Relationship between polarization information and full width at half maximum (FWHM). Gaussian polarization profiles were constructed for a range of standard deviation values, and then polarization information (bits) and FWHM (degrees) were calculated for each and plotted ($n_b=512$). B: Relationship between total information and signal-to-noise ratio. Total information was computed using the Gaussian channel approximation for $n_b = 64$ bins/receptors (solid). In addition, total information was approximated by the expression $\frac{\log e}{4} \frac{g^2 r^2 n_b}{N}$ (dashed) in which $g$ is the gradient slope, $r$ is the disk radius, and $N$ is the noise variance. Also shown is the polarization information ($0.8 \times I_{tot}$, dotted). The signal-to-noise ratio is expressed as $\frac{g}{\sqrt{N}}$. C: Relationship between projection directional accuracy ($\cos(\theta)$) and signal-to-noise ratio ($\frac{g}{\sqrt{N}}$). The Blahut-Arimoto algorithm was used to convert the polarization information values from 2B into directional accuracy as measured by the cosine of the angle $\theta$ between projection and gradient directions. The dotted line indicates $\cos(\theta)=1$.}
\label{fig2}
\end{figure}

Here we define polarization magnitude as $S_u - S_p$ which is the decrease in polarization entropy from the unpolarized to the polarized state, and as shown above is equal to the polarization information, $I_p$. This measure does not depend on the bin size (i.e. number of bins) as long as the bin size is smaller than the polarization width. In Fig~\ref{fig2}A, we plot a comparison of FWHM with polarization magnitude for a series of Gaussian polarization profiles of varying width $\sigma$. One can observe a narrower polarization width (smaller FWHM) corresponds to increased information (bits). In the context of yeast mating, the positioning of a punctate polarisome ($\text{FWHM} \sim 15^o$) in the correct location requires a certain amount of information ($\sim 3.77$ bits) gleaned from the pheromone gradient.

\subsection{Information capacity of gradient sensing from Gaussian channel approximation}

We are interested in a cell acquiring polarization information from a chemical gradient. What is the maximum rate at which the gradient transmits information about its direction to the polarizing cell? The answer can tell us about how fast a cell can polarize in a directionally accurate fashion. In spatial sensing, the cell collects information about the gradient by measuring the chemical concentration at different points on the cell surface. We represent the cell as a disk with individual receptors located on the external surface. Let $\tilde{c_i}$ be the local concentration of the gradient chemical at each receptor $i$. Because we are interested in the gradient direction, we normalize this concentration to the gradient midpoint for the cell: $c_i = \tilde{c_i} - c_{mid}$. 

Each receptor monitors the local concentration which fluctuates around a mean value. We approximate this noise as Gaussian~\cite{tostevin_mutual_2010,hu_quantifying_2011} so that the estimate $\hat{c_i} = c_i + z_i$ where $z$ is Gaussian with mean 0. In this manner, we model each receptor as a Gaussian channel with additive Gaussian noise. From information theory~\cite{cover_elements_2006}, the information capacity of the channel can be written as 
\begin{equation}
\label{eq:gauss}
I^{max} = \frac{1}{2}\log\left(1 + \frac{P}{N}\right)
\end{equation}
bits per second with the power constraint $P$ on the signal and noise variance $N$.

Every receptor represents an independent Gaussian channel, and so the power constraint can be written as a single term, the square of the ligand concentration sensed by the receptor $i$: $c_i^2 = P_i$. Thus, we can calculate $P_i$ from the gradient slope, and in the next section we describe how to estimate $N$, which together give us $I_i^{max}$. By summing the information over all receptors, we obtain the total information capacity of the cell: $I_{tot} = \sum_i I^{max}_i$

\subsubsection{Signal-to-noise perspective of Gaussian channel approximation}
When $N \gg P$, then $I^{max} = \frac{1}{2}\log(1 + \frac{P}{N}) \sim \frac{P \log e}{2N} \sim 0.72 \frac{P}{N} $. From a signal-to-noise perspective, one can view the information capacity as proportional to $\frac{P}{N}$. The signal $P$ in turn is related to the gradient slope since $P_i = c_i^2 = g^2r^2\cos^2(\theta_i)$ in which $g$ is the gradient slope, $r$ is the radius of the disk, and $\theta_i$ is the angle between the gradient direction and the positional vector of receptor $i$. Then, $P_{tot}= \sum_i P_i = g^2r^2\sum_i \cos^2(\theta_i)$.

For $n_b=64$ as an example, we plot (Fig~\ref{fig2}B) the information capacity as a function of $\frac{g}{\sqrt{N}}$, which is proportional to $\sqrt{\frac{P}{N}}$. As expected, as this signal-to-noise ratio increases, the information capacity increases, and for low signal-to-noise the information capacity is well approximated by $\frac{\log e}{2} \frac{P}{N} = \frac{\log e}{4} \frac{g^2 r^2 n_b}{N}$. In this regime, $I_{tot}$ increases linearly as a function of the slope squared, the radius squared, or the number of bins (S1~Fig). 

Using the Blahut-Arimoto algorithm as outlined by Andrews and Yglesias~\cite{andrews_information-theoretic_2007}, we converted the information values from Fig~\ref{fig2}B into polarity directional accuracy $\cos(\theta)$ values. The resulting plot (Fig~\ref{fig2}C) shows the relationship between $\cos(\theta)$ and $\frac{g}{\sqrt{N}}$ in this example. In this manner, one can identify the signal-to-noise range over which the directional accuracy of polarization is above a certain threshold. So all that remains to calculate $I_{tot}$ for real cells is to estimate the measurement noise $N$ at each receptor.

\subsection{Modeling ligand diffusion and receptor binding as a mixed Poisson distribution}

Under given biological conditions, what is the estimate of $N$, the measurement noise of a single receptor? As described in the Introduction, previous work has derived estimates of $N$ using various methods. The noise expression contains two terms: the variance from diffusion and the variance from stochastic receptor-ligand binding. However, there are subtle differences depending on the derivation. Thus, we adopted a new approach based on the Poisson mixture distribution~\cite{willmot_mixed_1986,karlis_mixed_2005,neyman_new_1939} to offer a new perspective and compare with previous work.  

A single receptor on a patch of cell surface membrane measures the concentration of a ligand in its local neighborhood via binding to the ligand (Fig~\ref{fig1}C). This measurement is corrupted by noise, which we wish to estimate. The binding process consists of two stochastic stages: 1) diffusion of ligand in/out of the local neighborhood of the receptor, and 2) binding/unbinding to receptor. We adopt the assumption~\cite{ten_wolde_fundamental_2016} that the local neighborhood is well-mixed with the global solvent environment. We define the neighborhood as a sphere of radius $s$ around the receptor.

We sketch the derivation here, with a more detailed description in the Supporting Information (S1~Text). 
Let $A_b = k_a c T(1-p)$ represent the number of binding events, and $A_u$, which equals $A_b$ at steady-state, represent the number of unbinding events in a time interval $T$ for a receptor. From chemical kinetics, $k_a$ is the association rate constant, $c$ is the concentration of ligand, and $p$ is the receptor occupancy.

Using the rule of error propagation~\cite{taylor_introduction_1996}, we express the fractional variance of $c$ as: $\left(\frac{\delta c}{c}\right)^2 = \frac{1}{c^2}\left(\frac{\partial c}{dA_b}\right)^2 \text{Var}[A_b] + \frac{1}{c^2}\left(\frac{\partial c}{dA_u}\right)^2 \text{Var}[A_u] = \frac{1}{(k_a c T (1-p))^2} \left(\text{Var}[A_b] + \text{Var}[A_u] \right)$.
Ligand unbinding is a Poisson process possessing a variance of $\text{Var}[A_u] = k_a c T (1 - p)$.
Estimating $\text{Var}[A_b]$ is more complicated because binding depends on the local concentration of ligand in the receptor neighborhood which is subject to fluctuations from diffusion. We can think of binding as the result of two successive Poisson processes: diffusion followed by the binding event.

In such a mixed Poisson distribution, the random variable $X$ is Poisson distributed while the rate parameter $\lambda$ is also a random variable so that $\text{Var}[X] = E[\lambda] + \text{Var}[\lambda]$.
If we let $c_l$ be the concentration of ligand in the local neighborhood of receptor and $c$ be the surrounding concentration, then the Poisson parameter for $A_b$ is $\lambda = k_a c_l T (1-p)$.
Given that $E[\lambda] = k_a c T (1-p)$ and $\text{Var}[\lambda] = (k_a T (1-p))^2 \text{Var}[c_l] $, we can write $ \text{Var}[A_b] = k_a c T (1-p) + (k_a T (1-p))^2 \text{Var}[c_l]  $.

The variable $c_l$ depends on the diffusive flux into ($f_i$) and out of ($f_o$) the local receptor neighborhood. Once again applying the rule of error propagation, we obtain $\text{Var}[c_l] = \left(\frac{\partial c_l}{\partial f_{i}}\right)^2 \text{Var}[f_{i}] + \left(\frac{\partial c_l}{\partial f_{o}}\right)^2 \text{Var}[f_{o}]$.
For a given receptor at steady-state, the flux out (and flux in) from the receptor neighborhood of radius $s$ is $f_o = 4\pi s Dc_lT = f_i$ molecules/s.
We thus have $ \text{Var}[f_i] = \text{Var}[f_o] = 4\pi s D cT(1-p) $.
We can now write
$ \text{Var}[A_b] = k_a c T (1-p) + \frac{2c(k_a T (1-p))^2}{4\pi s DT(1-p)}$.

Finally, putting it together,
\begin{equation}
\label{eq:noise}
\begin{split}
\left(\frac{\delta c}{c}\right)^2 &= \left(\frac{1}{(k_a c T(1-p))^2}\right) \left(k_a c T (1-p) +  \frac{2c(k_a T (1-p))^2}{4\pi s DT(1-p)} + k_a c T (1−p)\right) \\
&= \frac{2}{4\pi s DcT(1-p)} + \frac{2}{k_a c T(1-p)}.
\end{split}
\end{equation}
This expression is identical to that of Kaizu et al.~\cite{kaizu_berg-purcell_2014} and differs from Bialek and Setayeshgar~\cite{bialek_physical_2005} by a factor of $2(1-p)$ in the diffusion term.

\subsection{Numerical calculation of directional information}

The chemical gradient concentrations measured at the cell surface provide information about both the direction and slope (magnitude) of the gradient. The maximum information capacity captures both aspects of the gradient vector, whereas the polarization information ($I_p$) refers only to direction. What fraction of the total information capacity is directional?

We used numerical simulations to calculate the directional component of the gradient information. For a circular cell containing $n_b$ bins (receptors), we determined the average ligand concentration at each compartment $c_i$ based on the gradient and the radial location of the compartment. This concentration was corrupted by Gaussian noise of variance $N$.

As shown by Hu et al. $Z = Z_1 + jZ_2 = \sum_i c_i(\cos(\theta_i) + j\sin(\theta_i))$ is a sufficient statistic of the gradient direction ($\theta_i$ is angle between gradient vector and vector to bin $i$), and $\arctan(Z_2/Z_1)$ is an unbiased estimator. Thus, $Z$ contains full information about the gradient direction as determined from the ligand concentrations at the various receptors. For a given slope $g$ and noise variance $N$ (i.e. signal-to-noise ratio $g/\sqrt{N}$), we computed numerous $Z$ vectors which were used to create a polarization profile after discretization (see Methods). Then we determined the entropy and information ($I_p$) of the polarization profile as described in the first section (Eqs~\ref{eq:entropy} and \ref{eq:info}). This numerical calculation was compared to the theoretical information limit from the Gaussian channel approximation.

\begin{figure}[!ht]
\includegraphics[width=1.0\textwidth]{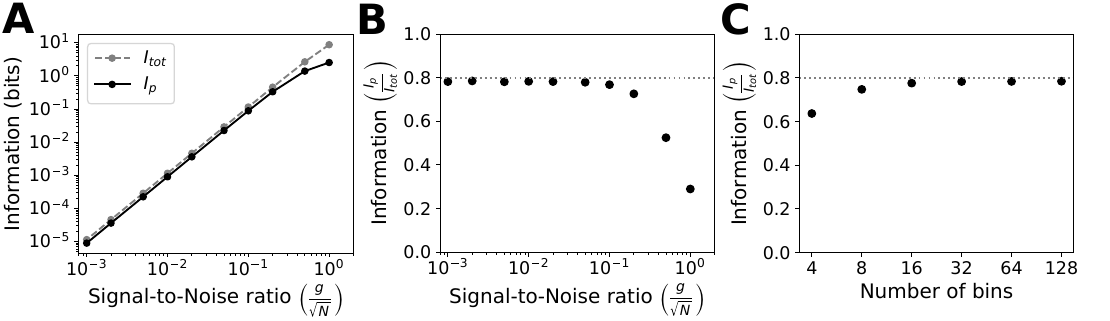} 
\caption{{\bf Estimating the directional component of total information from gradient.}
A: Numerical calculation of polarization information ($I_p$) compared to theoretical calculation of total information ($I_{tot}$). The directional component of total information was computed for a circular cell containing $n_b = 32$ bins subjected to a gradient of slope $g$ corrupted by noise $N$ for a range of signal-to-noise ratios ($\frac{g}{\sqrt{N}}$). It was compared to the total information, which  was determined using the Gaussian channel approximation summed over 32 bins. B: The data from 3A were replotted as the ratio of polarization information to total information ($\frac{I_p}{I_{tot}}$). Standard deviations are provided in S1~Table and are within the size of the data points on the plot. The dotted line is the ratio 0.8. C: The ratio of polarization information to total information as a function of the number of bins ($n_b$) for $\frac{g}{\sqrt{N}} = 0.01$.}
\label{fig3}
\end{figure}

In Fig~\ref{fig3}A, we estimated the directional information $I_p$ while varying the signal-to-noise (S/N) ratio for a fixed number of bins ($n_b = 32$). In Fig~\ref{fig3}B, the data are re-plotted as a proportion of the theoretical maximum information capacity $I_{tot}$. At high S/N ratios, the numerical values were far below the theoretical values because of discretization; the number of bins limited $I_p$ to a maximum of $\log n_b$, which in this case was 5 bits. At lower S/N ratios, we observed the polarization information converging to approximately 0.785 times the maximum capacity.   

In Fig~\ref{fig3}C, we investigated how discretization affected the maximum polarization information by varying the number of bins for a fixed signal-to-noise ratio ($\frac{g}{\sqrt{N}} = 0.01$). A smaller number of bins limited $I_p$ as described in the previous paragraph. For a larger number of bins, the polarization information once again converged to roughly 0.785 of the total information capacity.

We observed similar results for simulations of spherical cells with the directional information peaking at $\sim 0.785$ of the theoretical maximum (S2~Fig). In the Discussion, we outline an informal argument that the directional component of the gradient information should be $\sim 0.8 \times I_{tot}$. Below, we define the maximum polarization information by either numerical calculation or by multiplying the theoretical information capacity by 0.8.

\subsection{Using polarization information to evaluate cell polarity model simulations}

The information theory framework described above can be used to evaluate mathematical models of cell polarity. There are many such models ranging from models of specific biological systems (e.g. yeast pheromone-induced polarization) to more generic models that attempt to capture global features of cell polarization rather than specific mechanisms~\cite{levchenko_models_2002,mogilner_cell_2012}. How well do these models approach the theoretical limits? In particular, we assessed the extent and accuracy of polarization for a given signal-to-noise ratio.

As a test case, we employed three generic spatial models of cell polarization described in our previous work~\cite{chou_modeling_2008,chou_noise_2011}. The first model (Coop) consists of two partial differential equations with the first containing a cooperative ultrasensitive term for amplifying the external gradient (input $u$) to a steep internal gradient (i.e. polarization): $\frac{\partial a}{\partial t}=D \nabla_s^2 a+\frac{k_0}{1+(\beta u)^{-q}}-\left(k_2+k_3 b\right) a$. In this model, $a$ and $b$ are the two state variables ($a$ is the concentration of polarized species), and the second equation represents negative feedback regulation of the first equation (full model in S2~Text). The second model (PF) also consists of two equations but uses a positive feedback term containing the variable $a$ ($\frac{k_1}{1+(\gamma a)^{-h}}$) to provide the spatial amplification. The third model (filter-amplifier or FA) is composed of three equations in which the first equation ($ \frac{\partial f}{\partial t}=\frac{u-f}{\tau}$) is a first-order filter with time constant $\tau$, while the second and third equations are the positive feedback model.

\begin{figure}[!ht]
\includegraphics[width=1.0\textwidth]{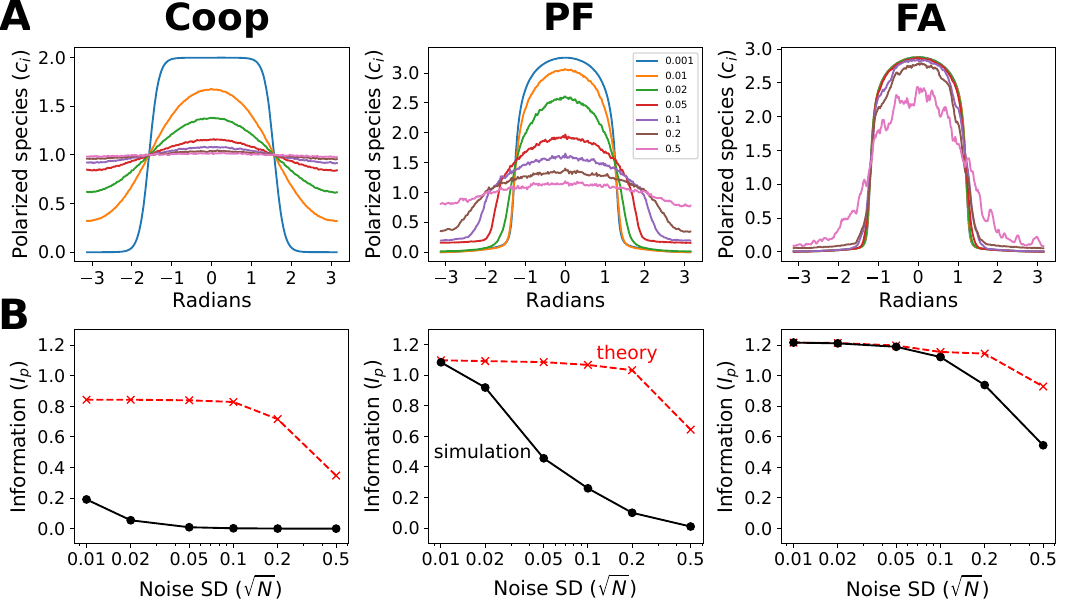} 
\caption{{\bf Comparing polarization information of simulations to theory.}
A: Polarization profiles at different noise standard deviation values ($\sqrt{N}$) for three models (from left to right): cooperative (Coop), positive feedback (PF), and filter-amplifier (FA). The polarization profiles on a 2D disk (polarized species in each bin marked in radians) represent the average peak polarization over 100 simulations for the following values of $\sqrt{N}$: 0.001, 0.01, 0.02, 0.05, 0.1, 0.2, 0.5. The gradient slope $g = 0.01\: \mu\text{m}^{-1}$
B: Polarization information ($I_p$) as a function of $\sqrt{N}$ for each of the three models. The simulation $I_p$ was computed from the average peak polarization profiles (solid black). There were three trials (each of 100 simulations); standard deviations are provided in S2~Table and are within the size of the data points on the plot. The theory $I_p$ was obtained from the convolution of 0-noise simulation with the numerically determined polarization profile (e.g. Fig~\ref{fig3}) for each noise value (red dashed).}
\label{fig4}
\end{figure}

We first examined the ability of the three models to polarize in a 1\% gradient in the absence of noise. For the 0-noise case, the theory states that the maximum information capacity is infinite so that $I_p$ can be the maximum possible (Eq~\ref{eq:gauss}). As shown in Fig~\ref{fig4}A, each of the models polarized significantly, amplifying the 1\% input gradient to produce a steeper output response, but each fell far short of the maximum achievable polarization in which the polarized species is concentrated in a single bin ($n_b = 400$) pointing in the correct direction. The polarization information ($I_p$) of the three models were 0.84, 1.1, and 1.2 bits, respectively, compared to the maximum $I_p = \log 400 = 8.64 \text{ bits}$ with a polarization peak value of 400.

A second challenge was assessing the noise-resistance of the models as increasing amounts of noise were added to the input gradient signal. One expects the average polarization over multiple simulations to degrade from noise either directly inhibiting the polarization magnitude or causing inaccurate polarization. Averaging a large number of simulations resulted in a profile that pointed in the correct direction, but the theory placed a bound on the maximum achievable polarization for a given signal-to-noise ratio. In addition, Eq~\ref{eq:noise} identifies one critical factor in this process which is the integration time of the model. Because it is in the denominator, integration time decreases the impact of ligand-binding noise by effectively averaging out stochastic fluctuations. All things being equal, a slower model should perform better in noise than a faster model. 

The effect of noise was measured relative to the 0-noise simulation, which represented the maximum polarization of the model for the input. Noise was added as additive Gaussian white noise to the 1\% gradient. For each noise value, 100 simulations were performed and averaged to obtain the output polarization profile. We then calculated the polarization entropy and information of this average polarization profile.

Qualitatively, we observed the loss of average polarization caused by increasing levels of noise in Fig~\ref{fig4}A. As described previously~\cite{chou_noise_2011}, the positive-feedback (PF) model was more noise resistant than the cooperative (Coop) model, while the two stage filter-amplifier (FA) model performed the best. This qualitative comparison, however, omits important quantitative considerations such as the speed of each model.

We assessed the integration time for the models by measuring the time-to-peak polarization after the input gradient was applied. In the 0-noise case the integration times were 0.79 (Coop), 2.0 (PF), and 5.2 (FA) seconds.

In Fig~\ref{fig4}B, we performed a more quantitative analysis by adjusting for integration times and comparing information values (see Methods). For a range of signal-to-noise (S/N) values, we observed the decrease in polarization information ($I_p$) with increasing noise (solid black line). To provide reference, we also calculated the maximum polarization information at each S/N ratio (dashed red line). This was calculated by computing the polarization profile for each S/N ratio as described in the previous section, and then convolving this profile with the 0-noise simulation polarization for a particular model. The convolution takes into account the fact that the 0-noise polarization is not perfectly amplified to a single bin, but spread over multiple bins.

The FA model exhibited the best noise-resistance, followed by the PF model, and then the Coop model was the worst. At low S/N, the FA model achieved close to the theoretical maximum, but there was some degradation of performance at higher levels of noise. The benefits of the filter-amplifier structure were best seen by comparing PF to FA; FA showed much better noise robustness while being roughly two-fold slower to reach peak polarization for this input gradient.

\subsection{Comparing theory to yeast experiments measuring polarization directional accuracy}

\begin{figure}[!h]
\includegraphics[width=1.0\textwidth]{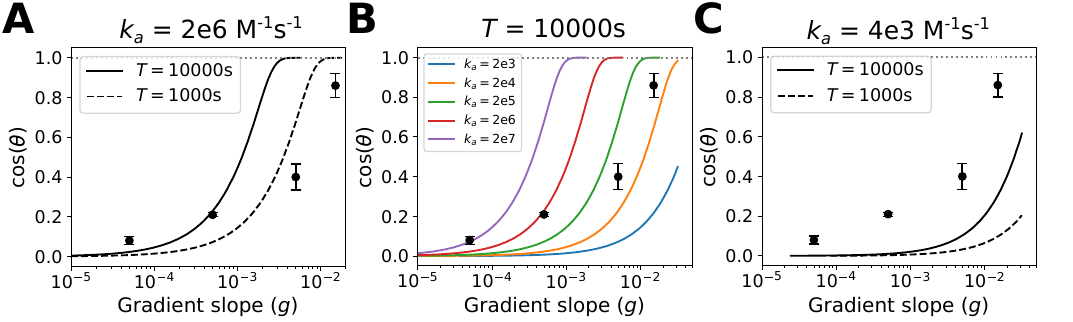} 
\caption{{\bf Comparing theory with yeast gradient sensing experiments.}
Polarization directional accuracy measured as $\cos(\theta)$ is plotted versus gradient slope ($\mu\text{m}^{-1}$). 
A: The theory curves represent the maximum polarization information calculated using Eq~\ref{eq:gauss} (multiplied by 0.8) converted to $cos(\theta)$ values. The noise value $N$ was estimated using  default parameter values (S3~Table) in Eq~\ref{eq:noise} with the exception of the integration time: $T_{lo} = 1000$s (dashed) and $T_{hi} = 10000$s (solid). The experimental data are from reference~\cite{chou_noise_2011}. B: The theory curves were generated over a ten thousand-fold range of $k_a$ (from $2 \times 10^3$ to $2 \times 10^7$ $\text{M}^{-1}\text{s}^{-1}$ as shown in the figure legend), and using the integration time $T_{hi} = 10000$s and the default values for the other parameters. C: The theory curves were generated using the lowest experimentally observed association rate constant $k_a$ from reference~\cite{ventura_utilization_2014}, $4 \times 10^3$ $\text{M}^{-1}\text{s}^{-1}$, and either $T_{lo} = 1000$s (dashed) or $T_{hi} = 10000$s (solid).}
\label{fig5}
\end{figure}

Haploid budding yeast cells project up a spatial gradient of mating pheromone toward the source, a mating partner~\cite{dohlman_regulation_2001,merlini_mate_2013,bardwell_walk-through_2005}. This directional growth of the mating projection occurs through a combination of gradient sensing and polarization response. In previous work, we measured the projection accuracy of \textit{S. cerevisiae} \textbf{a}-cells placed in alpha-factor pheromone gradients~\cite{moore_robust_2008,moore_yeast_2013,chou_noise_2011}. The published experiments were performed in microfluidics chambers which generated static spatial gradients with a midpoint concentration of 20~nM and slopes that varied from $0.005\%$ to $1.5\%$ per $\mu$m. Projection accuracy was quantitated in terms of $\cos(\theta)$ in which $\theta$ is the angle between the projection and gradient directions ($\cos(\theta) = 1$ is perfect polarization accuracy). The data are reproduced in Fig~\ref{fig5}A as $\cos(\theta)$ versus gradient slope for 4 slopes.

Using the theoretical framework outlined above (Eqs~1 to 4), we compared the yeast polarization accuracy data to the theoretical limits. The yeast cell was represented as a sphere of radius = 2 $\mu$m and possessing 10,000 receptors (bins) on the cell surface~\cite{yi_quantitative_2003}. According to the Gaussian channel approximation (Eq~\ref{eq:gauss}), we determined the maximum information per second transmitted by a gradient of slope $g$ and noise variance $N$. Multiplying this number by 0.8 (as described in the numerical estimation section) gave the maximum polarization information, $I_p$.

The noise was calculated according to Eq~\ref{eq:noise} with the free parameters being $D$ (diffusion constant), $k_a$ (receptor-ligand association rate constant), $T$ (integration time), and $s$ (radius of receptor neighborhood). The default values of the parameters are provided in S3~Table along with rough lower and and upper bounds. Plugging in these default values yielded $6.6 \times 10^{-6}$ for the diffusion noise, and $2.5 \times 10^{-2}$ for the binding noise. So for this particular example, binding noise dominates diffusion noise, and as a result, we focused on $T$ and $k_a$. 

To make a direct comparison to the data points, we converted the information values into the corresponding $\cos(\theta)$ directional accuracy measure using rate distortion theory. More specifically, we followed the procedure outlined by Andrews and Yglesias~\cite{andrews_information-theoretic_2007} employing the Blahut-Arimoto algorithm to generate the rate distortion curve relating information to $\cos(\theta)$ (see Methods for details).

In Fig~\ref{fig5}A, we calculated and plotted the maximum theoretical $\cos(\theta)$ curves as a function of gradient slope for the default parameter values in the noise equation. Because of the uncertainty in the integration time $T$, we explored a lower and higher value representing the approximate lower and upper bounds. When compared to the experiments, we observed that at shallower slopes the data points fell near the theoretical bound for the longer integration time (10000s), whereas at steeper gradient slopes the data points were under the theoretical bounds for both the shorter (1000s) and longer integration times. One would expect the data points to lie below the theoretical curves, which represent the maximum gradient sensing performance. Increasing integration time shifted the curve to the left.

Interestingly, the experimental literature exhibits a sizable discrepancy in measured values for the alpha-factor receptor association rate constant $k_a$ from $4 \times 10^3$ to $2 \times 10^6$ $\text{M}^{-1}\text{s}^{-1}$ (S4~Table). In Fig~\ref{fig5}B, we explored a range of values over 4 orders of magnitude, from $2 \times 10^3$ to $2 \times 10^7$ $\text{M}^{-1}\text{s}^{-1}$, with the other parameters set at their default values ($T = 10000s$). None of the curves fit all of the data points well. Larger $k_a$ values fit the shallower two points better, whereas the slower $k_a$ values produced curves that were closer to the steeper gradient data (i.e. experimental data are close to maximum). Only the fastest $k_a$ resulted in a theoretical maximum curve near or above all 4 data points. Increasing $k_a$ shifted the curve to the left.

Finally investigating the slowest of the experimentally measured $k_a$ values~\cite{ventura_utilization_2014}, we plotted, the theoretical curves for the slow and fast integration times for $k_a = 4 \times 10^3$ $\text{M}^{-1}\text{s}^{-1}$ (Fig~\ref{fig5}C). For both of the curves, all of the data points lie above the theoretical maximum values indicating they are not consistent with the theory using these particular values of $k_a$ and $T$.

\section{Discussion}

In this work we characterized spatial gradient sensing and polarization response from an information perspective. Applying an information theoretic framework that is a synthesis of previous work in the literature~\cite{ten_wolde_fundamental_2016,andrews_information-theoretic_2007,hu_quantifying_2011}, we interpreted cell polarity as representing the information the cell possesses about gradient direction, which can be defined mathematically as the entropy of uniform polarization minus the entropy of the polarized state (Eq~\ref{eq:info}). We estimated the theoretical limit of spatial sensing and response by modeling each receptor on the cell surface as a Gaussian channel, and then summing over all receptors. The signal is the chemical concentration at a given receptor which is proportional to the gradient slope, and the noise arises from ligand diffusion and receptor-binding. The noise terms could be modeled as a mixed Poisson distribution allowing us to derive an expression for the sensing noise of a single receptor that agrees with ten Wolde and colleagues~\cite{kaizu_berg-purcell_2014,ten_wolde_fundamental_2016}, while being slightly different from the results of fluctuation-dissipation theory~\cite{bialek_physical_2005}. The theory only models external noise assuming internal noise is 0; non-zero internal noise would further impair directional polarization.

The gradient vector possesses a direction and magnitude, and using numerical calculations, we were able to estimate the directional component of the total information. As the signal-to-noise ratio decreases, the polarization information also decreases, but the directional component approaches a limit of approximately 0.8 of the total information (Fig~\ref{fig3}). We offer the following heuristic justification. The possible gradient direction is uniformly distributed between 0 and $2\pi$ with differential entropy $\ln(2\pi)$. Taking into account the gradient magnitude (slope), the total differential entropy is $\ln(2\pi) + \ln(\frac{\pi}{2})$ in which the second term represents the differential entropy of the azimuthal angle direction. We approximate the directional fraction of the information as the directional entropy divided by the full entropy, $\ln(2\pi)/(\ln(2\pi) + \ln(\frac{\pi}{2})) \sim 0.8$, which is close to our numerically-derived value.

One application of the theory is to evaluate the performance of model simulations of polarization. As an example, we compared three generic polarization models described in previous work~\cite{chou_modeling_2008,chou_noise_2011}. First we examined the extent of polarization measured using polarization information, and as expected, the generic models did not polarize perfectly in the 0-noise case, but they still polarized with $I_p >> 0$.

Then we investigated the robustness of the models to noise by comparing the simulations in the presence of noise to the corresponding theoretical limits. Increasing the amount of noise added to the gradient input suppressed the average polarization of the model simulations. Among the three models, we reproduced the trend observed previously~\cite{chou_noise_2011} that the filter-amplifier (FA) was more noise-resistant than the positive-feedback (PF) model which was  better than the cooperative (Coop) model. However, now we could quantify these differences which take into account the speed of each model so that the noise robustness is not due to slowing down the dynamics to allow for more noise averaging. Interestingly, the FA architecture performed close to the theoretical limit in the low noise regime but performance declined at higher noise levels.

A second application was comparing the theoretical bounds to experimental data on yeast polarization accuracy in a pheromone chemical gradient. Following the example of Andrews and Iglesias~\cite{andrews_information-theoretic_2007}, we used rate-distortion theory to convert polarization information into a standard directional accuracy metric ($\cos(\theta)$) that could be compared to experiments. An initial observation from the theory was that the binding noise dominated diffusion noise, and so we could focus our attention on the parameters that most affected the signal-to-noise ratio and hence accuracy. By exploring the receptor binding noise parameter space we were able to identify parameter regions that gave rise to theoretical curves that were approximately consistent with the data.

More specifically, the theory highlights the importance of integration time, the time interval over which receptors monitor ligand concentration, and the association rate of ligand binding to receptor. Both place strict limits on the ability of the cell to polarize accurately up a gradient as parameters in the denominator of the binding noise equation (Eq~\ref{eq:noise}). In the case of yeast, there is uncertainty on how to interpret the integration time. A lower bound for $T$ is the time it takes the yeast cell to polarize key polarity proteins such as Bem1 and Cdc24 which is roughly 1000s~\cite{pringle_establishment_1995}. An upper bound is the length of the experiments which was on the order of 10000s, but the cell may not be actively sensing and polarizing for the whole experiment. The intuition is that a longer integration time allows the cell to average the binding noise over a longer time interval.

For the association rate constant $k_a$ there is disagreement in the literature with the values falling between $4 \times 10^3$ to $2 \times 10^6$ $\text{M}^{-1}\text{s}^{-1}$ (S4~Table). The theory supports the higher $k_a$ values because the lower values are inconsistent with the experimental data which lie above the maximum accuracy predicted by the theoretical curves in Fig~\ref{fig5} for lower $k_a$. The comparison highlights the need for more quantitative yeast polarization experimental data to supplement the existing data. Regardless, the theoretical results underscore the importance of this particular rate constant. Temporal averaging (high $T$) can help compensate for a low $k_a$, but at the cost of slower mating dynamics. 

It should be noted that other G-protein coupled receptor (GPCR) gradient sensing systems possess much faster association rate constants. For example, $k_a$ has been measured to be $\sim 2 \times 10^7$ for the cAMP receptor in \textit{Dictyosteleum}~\cite{valkema_model_1994}, 2 to 8 $\times 10^7$ for the neutrophil fMLP receptor~\cite{hoffman_receptor_1996}, and $0.8 \times 10^7$ $\text{M}^{-1}\text{s}^{-1}$ for the CCR5 chemokine receptor~\cite{swinney_study_2014}. These values are roughly 10-fold faster than the fastest measured $k_a$ for yeast alpha-factor receptor.

From an evolutionary standpoint, one can ask why alpha-factor receptor does not possess a higher $k_a$? To maintain a similar $K_d$ (equilibrium dissociation constant), the dissociation rate constant $k_d$ would also have to be faster which may not be desirable. Another reason is that speed is not important to yeast mating, and so any excess noise from a slow $k_a$ can be attenuated by extended time averaging (i.e. longer integration time). There could be other processes that are rate-limiting such as maintaining cell wall integrity as the mating projection grows. A third possibility is that there is no need for optimal sensing and response performance from a signal-to-noise perspective as long as the signal is sufficiently large during mating, i.e. gradient is steep.

From the theory we also observed the dependence of polarization information on the number of receptors, which helps to explain the high density of alpha-factor receptors (10,000 per cell or approximately 200 per $\mu\text{m}^2$ of plasma membrane), as well as on the radius, i.e., the larger the cell, the better for gradient sensing. Finally, because receptor occupancy influences the noise with the term ($1-p$) in the denominator of Eq~\ref{eq:noise}, receptor saturation (average chemical concentration $>> K_d$) will result in more noise and reduced polarization information compared to lower receptor occupancy.

In the future, we plan to extend this framework from non-motile cell systems to gradient sensing motile systems like bacterial chemotaxis~\cite{berg_physics_1977,macnab_gradient-sensing_1972}. This generalization will involve synthesizing and extending existing concepts in the literature~\cite{sourjik_responding_2012,tindall_overview_2008}, and applying the results to model simulations and experimental data with the goal of evaluating overall biological performance, as well as further characterizing key parameters and tradeoffs of the systems.

\section{Methods}
\subsection{Numerical computation of polarization information}
A two-dimensional (2D) disk of $n_b$ bins and radius $r = 1$ was placed in a chemical gradient of slope $g$ and noise variance $N$. The concentration of ligand $c_i$ at each bin $i$ was $c_i = g\cos(\theta_i) + \omega_i$, in which $\theta_i$ is the angle between gradient vector and bin vector and $\omega_i \sim \mathcal{N}(0, N)$ is the Gaussian white noise term. A similar approach was used to determine the polarization information for a three-dimensional (3D) sphere summing over bins on the sphere surface. The Fibonacci method was used to distribute the bins uniformly over the sphere surface. The concentration of ligand at each bin was determined from the x-coordinate of the bin multiplied by the gradient slope.

From reference~\cite{hu_quantifying_2011}, we know that $z = a + jb = \sum_i c_i(\cos(\theta_i) + j\sin(\theta_i))$ is a sufficient statistic of the gradient direction, and $\arctan(\frac{b}{a})$ is an unbiased estimator. We ran this computation $10^9$ to $10^{10}$ iterations until convergence of the distribution $\frac{z}{\|z\|} \sim \mathcal{D}_z$. We then calculated the entropy of the distribution which was subtracted from $\log n_b$ to determine the polarization information: $I_p = \log n_b - S_{D_z}$. 

\subsection{Evaluating simulations of generic models}
The model simulations were performed using the model equations (main text and S2~Text) as described previously~\cite{chou_modeling_2008,chou_noise_2011}. Briefly, the cell was modeled as a 2D disk of radius $r=1$ with the cell surface being divided into 400 bins. Gaussian white noise of specified variance was introduced at each bin with a noise time step of $k_t = 0.01$s. Spatial derivatives were approximated using a second-order finite difference discretization. The temporal discretization was carried out using a fourth order Adams-Moulton predictor-corrector method with a time step of 0.0001s.

All model simulations were run with a $1\%$ gradient (0.01 $\mu\text{m}^{-1}$) while the square root of the noise was varied from 0 to 0.5 ($\sqrt{N} = [0.0, 0.001, 0.01, 0.02, 0.05, 0.1, 0.2, 0.5]$). For each noise value, we calculated the average polarization time course from 100 model simulations. The peak polarization profile was identified along with the time-to-peak (TTP). We calculated the entropy and polarization information (using Eq~\ref{eq:info}) of the peak profile. The polarization information of the 0-noise simulation represented the maximum polarization achievable by the model for the gradient input.

We assessed noise tolerance by comparing the polarization information of the model simulations to the theoretical maximum for a given noise level. We computed the maximum theoretical polarization profiles of each model as we varied the noise magnitude.
As described above, we calculated numerical polarization profiles for a range of noise values. These were then convolved with the 0-noise profile for each model, and then the entropy of the resulting convolved profile was computed. This produced a curve of entropy values versus noise. We calculated the adjusted noise value by dividing the noise by the model integration time. The entropy adjusted for the integration time was then interpolated from the entropy curve using the adjusted noise value. The theoretical maximum information at a given noise value was the maximum entropy for the model ($\log n_b$ bits) minus the adjusted entropy, and this was compared to the polarization information from the model simulations as shown in Fig~\ref{fig4}B. 

\subsection{Relationship between information and \texorpdfstring{$\cos(\theta)$}{cos(theta)} values}
Yeast directional polarization accuracy was calculated in terms of $\cos(\theta)$, and so we constructed a standard curve relating information values to $\cos(\theta)$ values. Rate distortion theory defines the rate distortion function to be the minimum amount of information that representation $\hat{X}$ provides about $X$ over all distributions $p(x,\hat{x})$ that satisfy the distortion constraint. In this case, the distortion is $\cos(\theta)$, and we want to know the minimum information needed to achieve a particular distortion constraint ($\cos(\theta)$ value). We used the Blahut-Arimoto algorithm as described in detail by Andrews and Yglesias~\cite{andrews_information-theoretic_2007} and Chapter 13 of reference~\cite{cover_elements_2006} to construct the curve used to convert information values into $\cos(\theta)$ values in Fig~\ref{fig5}.

\bibliography{CPT-1.bib}

\newpage

\appendix

\section{Appendix: Supporting information}

\paragraph*{S1 Fig.}
\label{S1_Fig}
{\bf  Relationship between information and number of bins or radius.} 

\paragraph*{S2 Fig.}
\label{S2_Fig}
{\bf Calculation of polarization information on three-dimensional sphere.}

\paragraph*{S1 Table.}
\label{S1_Table}
{\bf Data for Fig~3B and Fig~3C.} 

\paragraph*{S2 Table.}
\label{S2_Table}
{\bf Data for Fig~4B.}

\paragraph*{S3 Table.}
\label{S3_Table}
{\bf Parameters for noise calculation.}

\paragraph*{S4 Table.}
\label{S4_Table}
{\bf Experimental measurements of alpha-factor receptor association rate constant.}

\paragraph*{S1 Text.}
\label{S1_Text}
{\bf Derivation of expression for receptor measurement noise.}  

\paragraph*{S2 Text.}
\label{S2_Text}
{\bf Equations for Coop, PF, and FA models.}  

\newpage

\begin{figure}[!ht]
\includegraphics[width=1.0\textwidth]{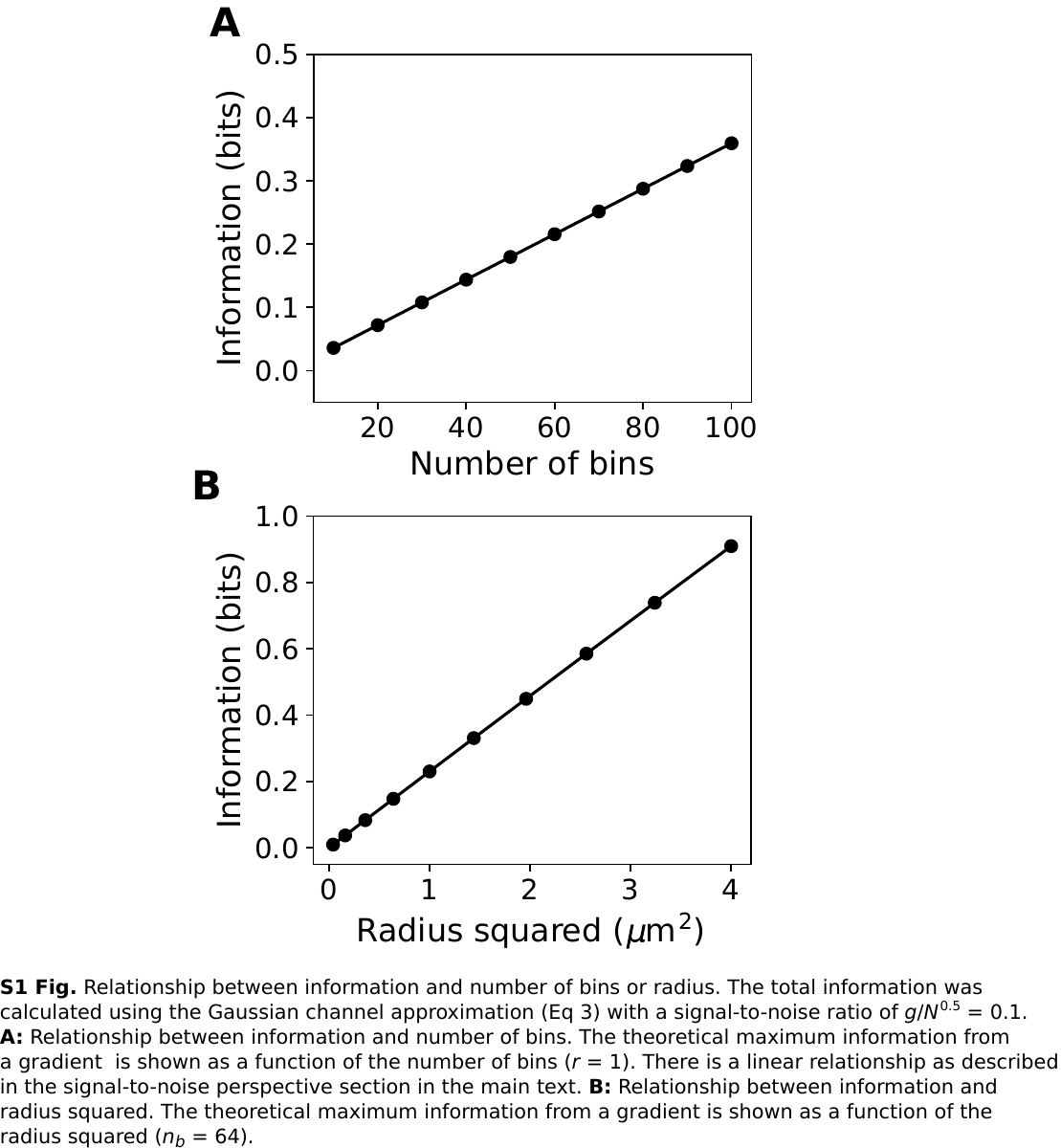} 
\end{figure}
\newpage

\begin{figure}[!ht]
\includegraphics[width=1.0\textwidth]{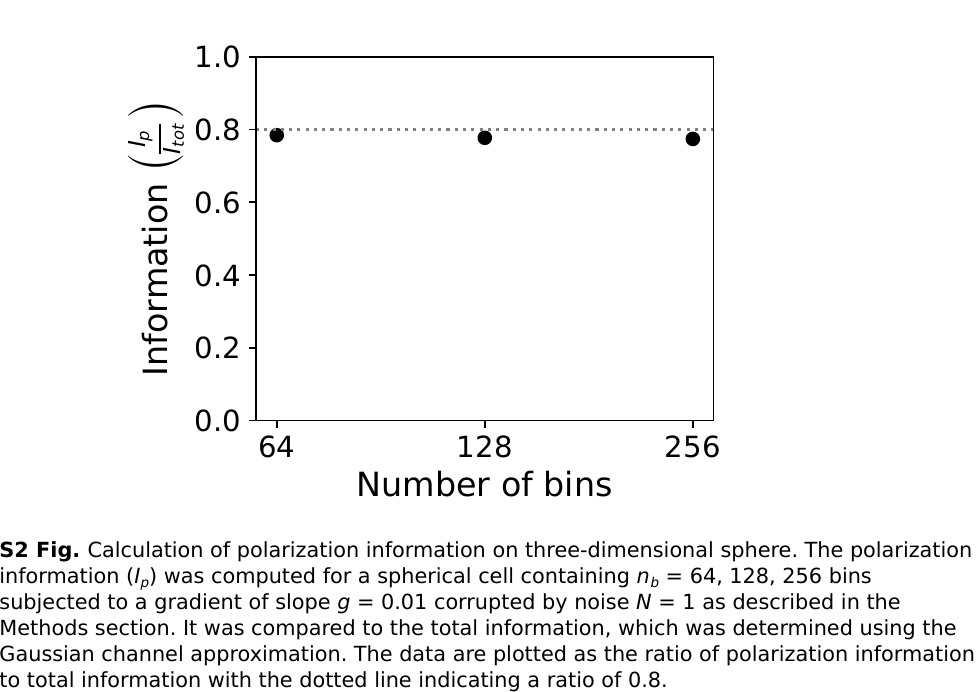} 
\end{figure}
\newpage

\begin{table}[!ht]
\caption*{{\bf S1~Table. Data for Fig~3B and Fig~3C.}}
\centering
\vspace{5pt}
\begin{tabular}{ |Sc||Sc|Sc| }
\hline
\multicolumn{3}{|c|}{\bf Fig 3B} \\ 
 \hline
 $g/\sqrt{N}$ & $I_p/I_{tot}$ & SD \\ 
 \hline
1 & 0.290 & $9.5 \times 10^{-6}$ \\ 
0.5 & 0.525 & $1.0 \times 10^{-5}$ \\ 
0.2 & 0.726 & $9.5 \times 10^{-5}$ \\ 
0.1 & 0.768 & $1.7 \times 10^{-4}$ \\ 
0.05 & 0.779 & $1.1 \times 10^{-4}$ \\ 
0.02 & 0.782 & $8.4 \times 10^{-4}$ \\ 
0.01 & 0.783 & $1.5 \times 10^{-3}$ \\ 
0.005 & 0.781 & $1.0 \times 10^{-3}$ \\ 
0.002 & 0.785 & $1.4 \times 10^{-3}$ \\ 
0.001 & 0.782 & $5.0 \times 10^{-3}$ \\ 
 \hline
 \hline
 \multicolumn{3}{|c|}{\bf Fig 3C} \\ 
 \hline
  $n_b$ & $I_p/I_{tot}$ & SD \\ 
 \hline
 4 & 0.637 & $3.2 \times 10^{-3}$ \\ 
 8 & 0.748 & $1.5 \times 10^{-3}$ \\ 
16 & 0.776 & $2.5 \times 10^{-3}$ \\ 
32 & 0.783 & $1.5 \times 10^{-3}$ \\ 
64 & 0.784 & $4.3 \times 10^{-4}$ \\ 
128 & 0.785 & $3.7 \times 10^{-4}$ \\ 
 \hline
\end{tabular}
\begin{center} Mean ($I_p/I_{tot}$) and standard deviation (SD) for 3 trials.
\end{center}

\label{table:S1_Table}
\end{table}
\newpage

\begin{table}[!ht]
\caption*{{\bf S2~Table. Data for Fig~4B.}}
\centering
\begin{tabular}{ |Sc||Sc|Sc|Sc|Sc|Sc|Sc| } 

\multicolumn{1}{c}{} & \multicolumn{2}{c}{\bf Coop} & \multicolumn{2}{c}{\bf PF} & \multicolumn{2}{c}{\bf FA}\\ 
\hline
 $\sqrt{N}$ & $I_p$ & SD & $I_p$ & SD & $I_p$ & SD \\ 
 \hline
0.01 & 0.19    & $3.2 \times 10^{-5}$ & 1.1  & $5.3 \times 10^{-3}$ & 1.2   & $8.4 \times 10^{-4}$ \\ 
0.02 & 0.055   & $3.2 \times 10^{-5}$ & 0.92 & $1.9 \times 10^{-2}$ & 1.2   & $1.2 \times 10^{-3}$\\ 
0.05 & 0.0091  & $2.3 \times 10^{-5}$ & 0.46 & $9.2 \times 10^{-3}$ & 1.2   & $1.6 \times 10^{-3}$\\ 
0.1  & 0.0023  & $2.2 \times 10^{-5}$ & 0.26 & $9.4 \times 10^{-4}$ & 1.1   & $8.4 \times 10^{-3}$\\ 
0.2  & 0.00058 & $6.0 \times 10^{-6}$ & 0.10 & $1.1 \times 10^{-3}$ & 0.94  & $1.3 \times 10^{-2}$\\ 
0.5  & 0.00010 & $4.2 \times 10^{-6}$ & 0.01 & $5.2 \times 10^{-4}$ & 0.54  & $2.7 \times 10^{-3}$\\ 
 \hline
\end{tabular}
\begin{center} Mean ($I_p$) and standard deviation (SD) for 3 trials.
\end{center}

\label{table:S2_Table}
\end{table}
\newpage

\begin{table}[!ht]
\caption*{{\bf S3~Table. Parameters for noise calculation.}}
\centering
\begin{tabular}{ |Sc||Sc|Sc|Sc|Sc| } 

\hline
Parameter & Description & Default & Bounds & Reference \\ 
 \hline
$r$ & cell radius  & 2 $\mu$m  & 2 - 4 $\mu$m & \cite{sherman_getting_2002} \\ 
$n_b$ & receptors per cell & 10,000    & 5000 - 20,000 & \cite{jenness_binding_1986,yi_quantitative_2003}\\ 
$D$ & $\alpha$-factor diffusion constant   & 100 $\mu\text{m}^2 \text{s}^{-1}$ & $\sim 100$ $\mu\text{m}^2 \text{s}^{-1}$ & \cite{young_estimation_1980,chen_modelling_2016}\\
$s$ & receptor neighborhood radius  & 0.01 $\mu$m    & 0.004 - 0.04 $\mu\text{m}$ & *\\
$c_0$ & mean $\alpha$-factor concentration & 20~nM  & 20~nM & \cite{chou_noise_2011} \\
$K_d$ & receptor dissociation constant  & 5~nM  & 5 - 10~nM & \cite{jenness_binding_1986,yi_quantitative_2003}\\
$k_a$ & receptor-ligand association rate  & $2 \times 10^{6}$ $\text{M}^{-1}\text{s}^{-1}$ & $4 \times 10^{3}$ - $2 \times 10^{6}$ & S4 Table \\
$T$ & integration time & 10000 s & $1000-10000$ s & \cite{chou_noise_2011, pringle_establishment_1995} \\

\hline
\end{tabular} \\
\vspace{6pt}
*Range spans receptor diameter to separation of receptors on cell surface.
\label{table:S3_Table}
\end{table}
\newpage

\begin{table}[!ht]
\caption*{{\bf S4~Table. Experimental measurements of alpha-factor receptor association rate constant.}}
\centering
\begin{tabular}{ |Sc|Sc| } 

\hline
Date [Reference] & $k_a$ ($\text{M}^{-1}\text{s}^{-1}$) \\ 
 \hline
1983~\cite{jenness_binding_1983}  & $3 \times 10^{3}$ \\ 
1986~\cite{jenness_binding_1986} & $3 \times 10^{5}$ \\
1988~\cite{raths_peptide_1988}   & $7 \times 10^{4}$ \\
2003~\cite{yi_quantitative_2003}   & $2 \times 10^{6}$ \\ 
2004~\cite{bajaj_fluorescent_2004}   & $2 \text{ - } 5 \times 10^{5}$ \\
2014~\cite{ventura_utilization_2014}   & $4 \times 10^{3}$ \\
 \hline
\end{tabular}

\label{table:S4_Table}
\end{table}
\newpage
\section*{S1 Text. Derivation of expression for receptor measurement noise.}

Under given biological conditions, what is the estimate of $N$, the measurement noise of a single receptor? As described in the Introduction, previous work has derived estimates of $N$ using various methods. The noise expression contains two terms: the variance from diffusion and the variance from stochastic receptor-ligand binding. However, there are subtle differences depending on the derivation. Thus, we adopted a new approach based on the Poisson mixture distribution~\cite{willmot_mixed_1986,karlis_mixed_2005,neyman_new_1939} to offer a new perspective and compare with previous work.  

A single receptor on a patch of cell surface membrane measures the concentration of a ligand in its local neighborhood via binding to the ligand (Fig~1C). This measurement is corrupted by noise, which we wish to estimate. The binding process consists of two stochastic stages: 1) diffusion of ligand in/out of the local neighborhood of the receptor, and 2) binding/unbinding to receptor. We adopt the assumption~\cite{ten_wolde_fundamental_2016} that the local neighborhood is well-mixed with the global solvent environment. We define the neighborhood as a sphere of radius $s$ around the receptor.
 
Let $A_b$ represent the number of binding events, and $A_u$ represent the number of unbinding events in a time interval $T$ for a receptor. From chemical kinetics, $A_b = k_a c T(1-p) $ where $k_a$ is the association rate constant, $c$ is the concentration of ligand, and $p$ is the receptor occupancy. The $(1-p)$ term represents unbound receptor that are capable of binding ligand. At steady-state, $A_b = k_a c (1-p) = k_d p = A_u $, where $k_d$ is the dissociation rate constant. Hence $c$ can be written in terms of $A_b$ and $A_u$.

Using the rule of error propagation~\cite{taylor_introduction_1996}, we express the fractional variance of $c$ as: $\left(\frac{\delta c}{c}\right)^2 = \frac{1}{c^2}\left(\frac{\partial c}{dA_b}\right)^2 \text{Var}[A_b] + \frac{1}{c^2}\left(\frac{\partial c}{dA_u}\right)^2 \text{Var}[A_u]$. We know that $ \frac{\partial c}{\partial A_b} = \frac{1}{k_aT(1-p)} = \frac{\partial c}{\partial A_u} $ so that $\left(\frac{\delta c}{c}\right)^2 = \frac{1}{(k_a c T (1-p))^2} \left(\text{Var}[A_b] + \text{Var}[A_u] \right)$.

Ligand unbinding is a Poisson process possessing an average of $k_d p T = k_a c T (1-p)$, and hence a variance of $ \text{Var}[A_u] = k_a c T (1-p) $.
Estimating $\text{Var}[A_b]$ is more complicated because binding depends on the local concentration of ligand in the receptor neighborhood which is subject to fluctuations from diffusion. We can think of binding as the result of two successive Poisson processes: diffusion followed by the binding event.

In a mixed Poisson distribution, the random variable is Poisson distributed while the rate parameter $\lambda$ is also a random variable so that the probability density $f(x)= \int_\Theta f(x|\lambda) g(\lambda) d\lambda $ where $\lambda \in \Theta$. Furthermore for the mixed Poisson distribution, $\text{Var}[X] = E[\lambda] + \text{Var}[\lambda]$.

If we let $c_l$ be the concentration of ligand in the local neighborhood of receptor and $c$ be the surrounding concentration, then for the random variable $A_b$ the Poisson parameter $\lambda = k_a c_l T (1-p)$.
We have $E[\lambda] = k_a c T (1-p)$, and $\text{Var}[\lambda] = (k_a T (1-p))^2 \text{Var}[c_l] $ so that $ \text{Var}[A_b] = k_a c T (1-p) + (k_a T (1-p))^2 \text{Var}[c_l]  $.

The variable $c_l$ depends on the diffusive flux into the local receptor neighborhood, $f_i$, and the diffusive flux out of the neighborhood, $f_o$. Once again applying the rule of error propagation, we obtain $\text{Var}[c_l] = \left(\frac{\partial c_l}{\partial f_{i}}\right)^2 \text{Var}[f_{i}] + \left(\frac{\partial c_l}{\partial f_{o}}\right)^2 \text{Var}[f_{o}]$.
For a given receptor, the flux out from diffusion over a time span $T$ from the receptor neighborhood of radius $s$ is $ f_o = 4\pi s Dc_lT$ molecules/s.

At steady-state, the average flux in equals the average flux out. Because this is a Poisson process, the variance is the same as the mean, which is equal to the inward average flux and variance.
We thus have $ \text{Var}[f_i] = \text{Var}[f_o] = 4\pi s D cT(1-p) $ with the factor $(1-p)$ representing the unoccupied receptors since we are estimating binding variance. Because $ \frac{\partial c_l}{\partial f_{i}} = \frac{1}{4\pi s DT(1-p)} = \frac{\partial c_l}{\partial f_{o}} $, $\text{Var}[c_l] = \left(\frac{2c}{4\pi s DT(1-p)}\right) $. We can now write
$ \text{Var}[A_b] = k_a c T (1-p) + \frac{2c(k_a T (1-p))^2}{4\pi s DT(1-p)}$.

Finally, putting it together,
\begin{equation}
\begin{split}
\left(\frac{\delta c}{c}\right)^2 &= \left(\frac{1}{(k_a c T(1-p))^2}\right) \left(k_a c T (1-p) +  \frac{2c(k_a T (1-p))^2}{4\pi s DT(1-p)} + k_a c T (1-p)\right) \\
&= \frac{2}{4\pi s DcT(1-p)} + \frac{2}{k_a c T(1-p)}. \nonumber
\end{split}
\end{equation}
This expression is identical to that of Kaizu et al.~\cite{kaizu_berg-purcell_2014} and differs from Bialek and Setayeshgar~\cite{bialek_physical_2005} by a factor of $2(1-p)$ in the diffusion term.
\newpage

\section*{S2 Text. Equations for Coop, PF, and FA models.}
The Coop and PF models are described in greater detail in Chou et al.~\cite{chou_modeling_2008,chou_noise_2011}. The FA model possesses a first-order filter in front of the PF model.
In all three models, $u$ is the input, $a$ is the polarized species on the surface of the 2D disk ($r=1$ $\mu$m), and $b$ is the negative feedback species that is well-mixed in all of the compartments. $D = 0.001$ $\mu\text{m}^2/s$ is the surface diffusion coefficient.
\subsection*{Cooperative Model (Coop)}

\begin{align*}
\frac{\partial a}{d t} &= D \nabla_s^2 a+\frac{k_0}{1+(\beta u)^{-q}}-k_2 a-k_3 b a \\
\frac{d b}{d t}&=k_4 \hat{a} b 
\end{align*}

\noindent where $\hat{a}=\bar{a}-a_{ss}$ and $\bar{a}=\frac{\int_s a d s}{\int_s d s}$. $k_0=10$, $k_2=k_3=k_4=1$, $\beta=1$, $q=1000$, $a_{ss}=1$.

\subsection*{Positive Feedback Model (PF)}

\begin{align*}
\frac{\partial a}{d t} &= D \nabla_s^2 a+\frac{k_0}{1+(\beta u)^{-q}}+\frac{k_1}{1+(\gamma a)^{-h}}-k_2 a-k_3 b a-k_5 \hat{a} \\
\frac{d b}{d t}&=k_4 \hat{a} b 
\end{align*}

\noindent where $\hat{a}=\overline{a}-a_{ss}$ and $\bar{a}=\frac{\int_s a d s}{\int_s d s}$. $k_0=1$, $k_1=10$, $k_2=k_3=k_4=1, k_5=10$, $\beta=1$, $\gamma=\frac{1}{1+u^{-q}}$, $q=100$, $h=2$, $a_{ss}=1$.

\subsection*{Filter-Amplifier Model (FA)}

 The FA model is composed of three equations in which the first equation is a first-order filter with time constant $\tau$, while the second and third equations are the positive feedback (PF) model.

 \begin{align*}
\frac{\partial f}{\partial t} &= \frac{u-f}{\tau} \\
\frac{\partial a}{d t} &= D \nabla_s^2 a+\frac{k_0}{1+(\beta f)^{-q}}+\frac{k_1}{1+(\gamma a)^{-h}}-k_2 a-k_3 b a-k_5 \hat{a} \\
\frac{d b}{d t}&=k_4 \hat{a} b 
\end{align*}

\noindent where $\tau=10$, and the other parameters are specified above in the PF model.

\end{document}